\documentclass[12pt]{iopart}

\newcommand{\eqb}{\begin{eqnarray}}
\newcommand{\eqe}{\end{eqnarray}}
\newcommand{\diff}{\textrm{d}}
\usepackage{graphicx}
\usepackage{pst-grad}

\begin{document}

\title[A current driven instability in relativistic shocks]
{A current driven instability in parallel, relativistic shocks}

\author{B. Reville$^{1\dagger}$, J.G. Kirk$^2$,
P. Duffy$^1$}

\address{
$^1$School of Mathematical Sciences, University College Dublin, Ireland\\
$^2$Max-Planck-Institut f\"ur Kernphysik, 69029 Heidelberg, Germany
}
\ead{$^\dagger$brian.reville@ucd.ie}
\begin{abstract}
Recently, Bell \cite{bell04} has reanalysed the problem of wave excitation by
cosmic rays propagating in the pre-cursor region of a supernova remnant shock
front. He pointed out a strong, non-resonant, 
current-driven instability that had been overlooked in the kinetic
treatments by Achterberg \cite{achterberg83} and 
McKenzie \& V\"olk \cite{mckenzievoelk82}, and suggested that
it is responsible for substantial amplification of the ambient magnetic field. 
Magnetic field amplification is also an important issue in the problem of the
formation and structure of relativistic shock fronts, particularly in relation
to models of gamma-ray bursts. 
We have therefore generalised the linear 
analysis to apply to this case, assuming 
a relativistic background plasma and a monoenergetic,
unidirectional incoming proton beam. We find essentially the same 
non-resonant instability noticed by Bell, and show that also 
under GRB conditions, it 
grows much faster than the
resonant waves.  We quantify the extent to which 
thermal effects in the background plasma limit 
the maximum growth rate.
\end{abstract}

\submitto{\PPCF}
\maketitle

\section{Introduction}
The acceleration of cosmic rays at the shock front that bounds a supernova
remnant is thought to proceed via the diffusive first-order Fermi mechanism
(for reviews, see \cite{blandfordeichler87,drury83,jonesellison91}). 
In the shock precursor, where the cosmic rays stream at roughly the shock
speed, Alfv\'en waves 
can grow as a result of interacting 
at the cosmic ray cyclotron resonance.
In turn, these waves provide the pitch-angle scattering essential for the
acceleration process \cite{mckenzievoelk82}. 
Using a kinetic approach, Achterberg \cite{achterberg83} 
found that the presence of cosmic rays has no significant impact on the nature 
of the plasma modes. 
However, Bell \cite{bell04}, using a hybrid MHD/kinetic theory
approach,
has recently challenged this
conclusion. He identifies a low frequency, 
non-resonant, electromagnetic mode that can be strongly driven by
the current induced in the plasma by the streaming cosmic rays. 
Simple analytic estimates and MHD simulations both suggest that the nonlinear
evolution of this instability leads to very strong 
amplification of the ambient magnetic field. In turn, this may result in 
a cosmic-ray acceleration rate that is much more rapid than previously 
thought \cite{bell05,belllucek}.

Gamma-ray bursts drive highly relativistic outflows with 
Lorentz factors of 100 or more \cite{grbreview}. On interacting with the 
surrounding medium, a shock front forms, but the mechanism by which this
happens is controversial \cite{spitkovsky05}. Observations of the afterglows of
gamma-ray bursts suggest that the ambient magnetic field must
be amplified substantially at the shock front. 
The relativistic Weibel
or two-stream instability, that generates a small-scale magnetic field
in a previously unmagnetized plasma,
has been investigated in this connection 
\cite{medvedevloeb99}, but appears to saturate 
at a relatively low amplitude 
\cite{lyubarskyeichler05,wiersmaachterberg04}.
In recent years, particle in cell (P.I.C.) simulations have been used in
studies of collisionless shocks \cite{spitkovsky05}, 
in particular with reference to the Weibel instability 
~\cite{hededalnishikawa,nishikawaetal,silvaetal}. The instability
that is presented in this paper has not been identified to date. 
However, relativistic P.I.C. electron-proton simulations in three  
dimensions still present a major
challenge, even to the best computational resources currently available.

When two electron-proton plasmas moving at high speed 
relative to each other collide and interpenetrate, the first process that
takes place is the equilibration of the electrons
\cite{lerchepohlschlickeiser}. However, 
this releases only a small fraction of the available free energy.   
The shock front responsible for the thermalisation of 
the bulk of the energy is mediated by the interaction of the two
counterstreaming proton plasmas.  Thus, 
Pohl \& Schlickeiser \cite{pohlschlickeiser} 
attacked the 
shock formation problem  
starting with a monoenergetic relativistic 
proton beam penetrating a relatively dense, cool
background in a direction aligned with the ambient magnetic field. They 
then 
computed the isotropisation rate that
results from the resonant excitation of Alfv\'en waves in the background
plasma. Here, we use the same initial conditions, but, 
because this physical situation resembles that in
the precursor region of a SNR shock front, we analyse
instead the
growth rate of the relativistic analogue of the non-resonant mode 
discovered by 
Bell \cite{bell04}. We find it grows much faster than the resonant
mode, and, as in the case of SNR, 
can be expected to generate a substantial 
magnetic field transverse to the beam direction.
However, we find the instability is quite sensitive to damping by thermal 
effects, once the background plasma is heated.

\section{Dispersion relation}
The linear dispersion relation for the propagation of transverse
waves parallel to the magnetic field in a plasma made up 
of components labelled by $j$ is:
\eqb
n_\|^2-1-\sum_j \chi_j(k,\omega)&=&0
\label{dispersionrel}
\eqe
where $n_\|=ck/\omega$ is the refractive index 
and $\chi_j(k,\omega)$ is the
susceptibility of the $j$'th component, for 
wavenumber $k$ and frequency $\omega$ ($>0$). 

The three components in the case we consider are:
\begin{enumerate}
\item
Protons that make up the background or downstream distribution, 
denoted by $j=\textrm{p}$. These have a number 
density $n_{\rm p}$ and a thermal
distribution with (dimensionless) 
temperature  $\Theta_{\rm p}=k_{\rm B}T_{\rm p}/m_{\rm p}c^2$.
The waves will be analysed in the 
lab.\ frame, chosen such that these particles have a vanishing net 
drift speed $c\beta_{\rm p}=0$. 
\item
Electrons that 
stem partly from the downstream plasma and partly from the incoming upstream 
plasma or beam. They also have a thermal distribution 
with lab.\ frame density $n_{\rm e}$ and temperature $\Theta_{\rm e}$. 
Their drift speed in the lab.\ frame is $c\beta_{\rm e}$ and  
their Lorentz factor $\Gamma_{\rm e}=\left(1-\beta_{\rm e}^2\right)^{-1/2}$.
\item
Protons of the upstream medium that form a monoenergetic, unidirectional 
incoming beam along the magnetic field. Their 
lab.\ frame density is $n_{\rm b}$ and their 
drift Lorenz factor $\Gamma_{\rm b}$. The beam 
distribution function is 
\eqb
\label{fbeam}
f_b(\vec{u}) = \left(\frac{1}{2 \pi u_\bot}\right)
\delta(u_\bot)\delta(u_\|-\Gamma_b\beta_b)
\eqe 
\end{enumerate}

Following Achterberg \cite{achterberg83}, we impose 
the conditions of overall charge
neutrality and zero net current on these components. 
This implies:
\eqb
\sum_j \omega_{{\rm p}j}^2/\omega_{{\rm c}j}&=&0
\label{chargefree}\\
\noalign{\hbox{and}}\nonumber
\sum_j \beta_j\omega_{{\rm p}j}^2/\omega_{{\rm c}j}&=&0
\label{currentfree}
\eqe
where $\omega_{{\rm p}j}=\left(4\pi n_j q_j^2/m_j\right)^{1/2}$, 
denotes the plasma 
frequency of the $j$'th component that consists of particles 
of charge $q_j$ and mass
$m_j$, 
and $\omega_{cj}=q_j B/m_jc$, denotes its cyclotron frequency. In the present
case $m_{\rm b}=m_{\rm p}$ is the proton mass 
and $q_{\rm b}=q_{\rm p}=-q_{\rm e}=e$ is the modulus of the 
electronic charge. 

For parallel propagation, the susceptibility $\chi_j$ is given 
by Yoon \cite{yoon90}. The quantity $\omega^2\chi_j$, which is invariant to
Lorentz boosts along the magnetic field, can be written:
\eqb
\omega^2\chi_j&=&
\omega_{{\rm p},j}^2\int{\diff^3 u\over\gamma} f_j(\vec{u})
\left[{-\omega\gamma+cku_\|\over D\left(u_\|\right)} 
- {u_\bot^2\over 2}{(c^2k^2-\omega^2)\over D^2\left(u_\|\right)}\right]
\label{yoon}
\eqe
Here, $f_j(\vec{u})$ 
is
the distribution function of particles of four velocity 
$c(\gamma,\gamma\vec{u})$, normalised such that $\int\diff^3u f_j=1$, the
components of $\vec{u}$ parallel and perpendicular to the magnetic field
direction are $u_\|$ and $u_\bot$, and the resonant denominator is
\eqb
D\left(u_\|\right)&=&\varepsilon\omega_{{\rm c}j}\left(1+
Z\left(u_\|\right)\right)
\label{resdenom}
\\
\noalign{\hbox{with}}
\nonumber\\
Z\left(u_\|\right)
&=&{\omega\gamma-cku_\|\over\varepsilon\omega_{{\rm c}j}}
\label{zsmall}
\eqe
and $\gamma=\sqrt{1+u_\bot^2+u_\|^2}$.
The waves are circularly
polarised, with $\varepsilon=1$ corresponding to left-handed, and 
$\varepsilon=-1$ to right-handed waves, for $k>0$.

The waves we consider are non-resonant for the electrons and background
protons. Furthermore, these components are \lq\lq magnetized\rq\rq,
in the sense that for all relevant values of $u_\|$, the 
resonant denominator  
defined in Eq.~(\ref{resdenom}) can be expanded using as a small
parameter $Z\left(u_\|\right)\ll1$:
\eqb
{1\over D^n\left(u_\|\right)}&\approx&
{1\over\left(\varepsilon\omega_{{\rm c}j}\right)^n}\left(
1-n Z\left(u_\|\right)+{n(n+1)\over 2!}Z^2\left(u_\|\right)+\dots\right)
\label{expansion}
\eqe
Inserting this expansion into Eq.~(\ref{yoon}), one finds, using 
$\int \diff^3u f_j\left(\vec{u}\right)u_\|/\gamma=\beta_j$, for
the zeroth $\chi^{(0)}$ and first order $\chi^{(1)}$ contributions:
\eqb
\omega^2\chi^{(0)}_j\left(k,\omega\right)&=&
{\omega_{{\rm p}j}^2
\over
\varepsilon\omega_{{\rm c}j}}
\left(c\beta_j k-\omega\right) -
{\omega_{{\rm p}j}^2
\over
2\omega_{{\rm c}j}^2}\left(c^2k^2-\omega^2\right)
\left<u_\bot^2/\gamma\right>_j
\label{zeroorder}\\
\omega^2\chi^{(1)}_j\left(k,\omega\right)&=&
{\omega_{{\rm p}j}^2
\over
\omega_{{\rm c}j}^2}
\left(
\omega^2\left<\gamma\right>_j 
-2c\omega k\left<u_\|\right>_j+c^2k^2\left<u_\|^2/\gamma\right>_j\right)
\nonumber\\
&&-
{\omega_{{\rm p}j}^2
\over
\varepsilon\omega_{{\rm c}j}^3}
\left(c^2k^2-\omega^2\right)
\left(
ck\left<u_\bot^2 u_\|/\gamma\right>_j-\omega\left<u_\bot^2\right>_j\right)
\eqe
where $\left<\dots\right>_j=\int\diff^3u\dots f_j\left(\vec{u}\right)$.

The susceptibilities $\chi_j$ describe the currents induced
by the wave field in each plasma component. The first term on the RHS of 
Eq.~(\ref{zeroorder}) is proportional to the charge and current
density of the $j$'th component in the unperturbed state. 
In a neutral, current-free plasma in which all components are
magnetized, i.e., 
can be treated using the expansion (\ref{expansion}), these terms cancel,
according to Eqs~(\ref{chargefree}) and (\ref{currentfree}). 
However, in the plasma we
consider, only the background protons and electrons are magnetized, 
and these terms do not cancel:
\eqb
\sum_{j=p,e}\omega^2\chi^{(0)}_j\left(k,\omega\right)=
{\omega_{{\rm pb}}^2
\over
\varepsilon\omega_{{\rm cb}}}
\left(\omega-c\beta_b k\right) -
\sum_{j=p,e}{\omega_{{\rm p}j}^2
\over
2\omega_{{\rm c}j}^2}\left(c^2k^2-\omega^2\right)
\left<u_\bot^2/\gamma\right>_j
\eqe
We describe the
waves in a reference frame in which the background protons have zero drift,
and assume they have an isotropic distribution in this frame. 
In this case $\left<u_\|\right>_{\rm p}=
\left<u_\bot^2u_\|/\gamma\right>_{\rm p}=0$ and 
$\left<u_\bot^2/\gamma\right>_{\rm p}=2\left<u_\|^2/\gamma\right>_{\rm p}$. 

Inserting the distribution function (\ref{fbeam}) into Eq.(\ref{yoon}),
we find the
susceptibility of the beam protons is 
\eqb
\omega^2\chi_{\rm b}(k,\omega)&=&
{ \omega_{\rm pb}^2\left(c\beta_{\rm b}k-\omega\right)
\over
\varepsilon\omega_{\rm cb} - \Gamma_{\rm b}
\left(c\beta_{\rm b}k-\omega\right)
}
\eqe
If 
$\Gamma_{\rm b}c\beta_{\rm b}\left|k\right|\ll\left|\omega_{\rm cb}\right|$,
the expansion 
(\ref{expansion}) can also be used for the beam protons, so that the 
plasma is fully compensated. However, 
modes of shorter wavelength, such that
\eqb
\left|\omega_{\rm cp,e}\right|/\Theta_{\rm p,e}&\gg& c|k| 
\,\gg\, \left|\omega_{{\rm cb}}\right|/\Gamma_{\rm b}
\eqe
are unable to induce a compensating current in the beam particles.
The overall susceptibility can then be written:
\eqb
\omega^2\chi&\approx&
{{\omega'_{\rm pb}}^2\omega'\over\epsilon\omega_{\rm c}} - 
{{\omega'_{\rm pb}}^2\omega'\over
\epsilon\omega_{\rm c}+\omega'}
+{c^2\omega^2\over v_{\rm A}^2}
+{\omega_{\rm pp}^2\omega\over\epsilon\omega_{\rm c}^3}
\left(c^2k^2-\omega^2\right)
\left<u_\bot^2\right>_{\rm p}
\label{chiapprox}
\eqe
where we have neglected the electron response, except for its contribution
to the overall current in the first term and to the 
quantity $v_{\rm A}$, defined as 
the corrected non-relativistic expression for the 
speed of an Alfv\'en wave in the background plasma~\cite{yoon90}:
\eqb
v_{\rm A}^2&=&c^2\left[\sum_{\rm p,e}{\omega_{{\rm p}j}^2\over
\omega_{{\rm c}j}^2}
\left(\left<\gamma\right>_j+\left<u_\bot^2/2\gamma\right>_j\right)\right]^{-1}
\eqe

The quantity
$\omega'=\Gamma_{\rm b}\left(\omega-\beta_{\rm b}ck\right)$ denotes the 
wave frequency as seen in the frame of the beam particles, and the plasma
frequency
of the beam particles in this frame is given by
${\omega'}_{\rm b}^2={\omega}_{\rm b}^2/\Gamma_{\rm b}$.

\begin{figure}
\hbox{%
\includegraphics[width=0.5\textwidth]{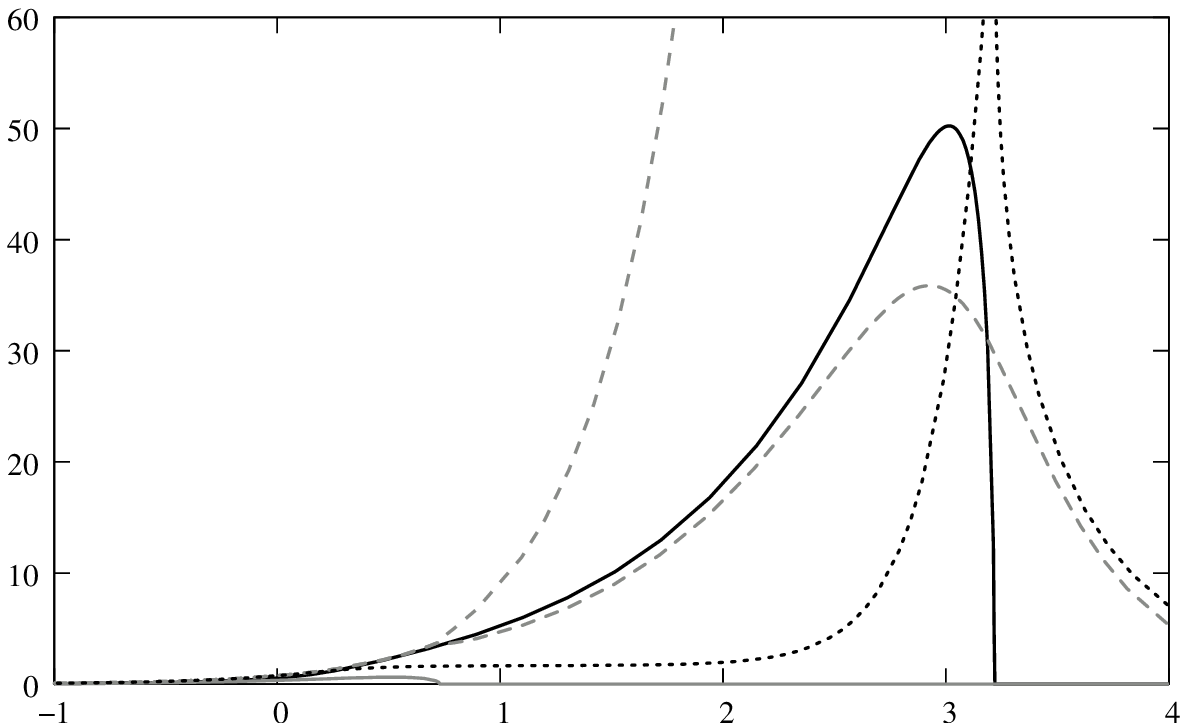}
\includegraphics[width=0.5\textwidth]{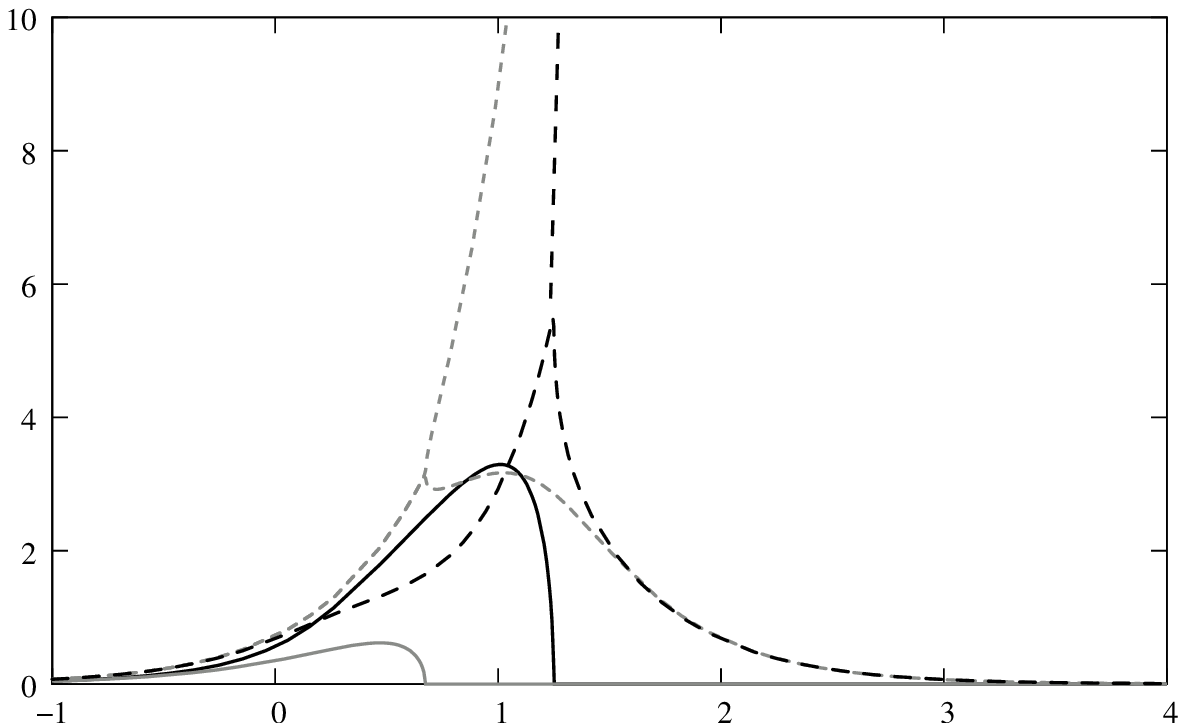}
}
\rput(7.1,0.5){$\log\tilde{k}$}
\rput(15.1,0.5){$\log\tilde{k}$}
\rput(0.1,4.8){$\tilde{\omega}$}
\rput(8.1,4.8){$\tilde{\omega}$}
\caption{%
\label{modeplot}%
Dispersion curves for 
the illustrative parameter set: $v_{\rm A}=2\times10^{-5}c$, $n_{\rm b}=n_{\rm
  p}/3$, $\Gamma_{\rm b}=10$ for two temperatures: 
$\Theta_{\rm p}=10^{-1}$ (right panel)
and $10^{-3}$ (left panel). 
The real part of the frequency of each mode is shown as a dashed 
line, the imaginary
part as a solid line. The left(right)-handed waves are shown in grey(black).}
\end{figure}

\section{Wave modes}
For a cold background
plasma, and for low frequency modes such that $\omega'\approx 
-\Gamma_{\rm b}\beta_{\rm b}ck$ with $\left|\omega'\right|\gg\left|\omega_{\rm cb}\right|
$, 
one recovers 
from Eq.~(\ref{chiapprox})
a simple form analogous to that discussed by
Bell \cite{bell04}:
\eqb
\omega^2&=&{v_{\rm A}^2\over c^2+v_{\rm A}^2}
\left( c^2k^2+{\beta_{\rm b}\omega_{\rm pb}^2ck\over \varepsilon\omega_{\rm
    cb}}
\right)
\label{bell1}
\eqe
which gives a purely growing mode ($\textrm{Re}(\omega)=0$) for the
right-handed polarisation $\varepsilon=-1$, provided the driving by the 
beam-induced current is sufficiently strong. 
For $v_{\rm A} \ll c $, the mode reaches a maximum growth rate 
\eqb
\textrm{Im}(\omega)& =& 
\frac{1}{2}\frac{n_{\rm b}}{n_{\rm p}}\beta_{\rm b}\omega_{\rm pp}
\label{bell2}
\eqe
which is independent of the magnetic field strength.
The corresponding wave number is 
\eqb
k_{\rm max} = \frac{1}{2}\frac{n_{\rm b}}{n_{\rm p}}\beta_{\rm b}
\frac{\omega_{\rm pp}}{v_{\rm A}}
\label{bell3}
\eqe 

The (non-resonant) thermal effects on this mode arising from the term
containing $\left< u_\bot^2\right>$ are easily analysed in the case of 
weak magnetic field $v_{\rm A}\ll c$, since then $\omega^2\ll c^2k^2$ and
$\omega'\approx -\Gamma_{\rm b}\beta_{\rm b}ck$. The 
dispersion relation is then approximately
\eqb
\tilde{\omega}^2&=& \frac{v_{\rm A}^2}{c^2} \tilde{k}\left(
\tilde{k}+{\Gamma_{\rm b}\beta_{\rm b}^2\omega_{\rm pb}^2
\over\varepsilon\omega_{\rm c}^2}-\tilde{k}{\left<u_\bot^2\right>
\omega_{\rm pp}^2\over\varepsilon\Gamma_{\rm b}\beta_{\rm b} \omega_{\rm c}^2} \tilde{\omega} \right)
\eqe
where dimensionless units are used 
in which $\tilde{k}=\Gamma_{\rm b}\beta_{\rm b}ck/\omega_{\rm c}$ and
$\tilde{\omega}=\Gamma_{\rm b}\beta_{\rm b}\omega/\omega_{\rm c}$.
The $\varepsilon=-1$ mode reaches a
maximum growth rate
of 
\eqb
\textrm{Im}(\omega)&=&\frac{\sqrt{3}}{2}\left(\frac{n_{\rm b}}{n_{\rm p}}\right)^{2\over3}
\left(\frac{v_{\rm A}}{c}\right)^{2\over3} \left(\frac{\omega_{\rm pp}}{\omega_{\rm c}}\right)^{2\over3}
\left(\frac{\beta_{\rm b}^2}{\left< u_\bot^2\right>}\right)^{1\over3}\omega_{\rm c}
\label{thermaleffects}
\eqe
In terms of the dimensionless temperature $\Theta_{\rm p}$ of the background
protons for the appropriately normalised J\"uttner-Synge distribution
\eqb
\left<u_\bot^2\right>&=&16\Theta_{\rm p}^2\left[1+3(1+\Theta_{\rm p})
\right]\textrm{e}^{-1/\Theta_{\rm p}}/
\left[K_2\left(1/\Theta_{\rm p}\right)\right]
\eqe
so that the growth rate given by Eq.~(\ref{thermaleffects}) falls off rapidly
with increasing temperature.

After multiplication by the factor 
$\epsilon\omega_{\rm c}-\omega'$, Eq.~(\ref{dispersionrel}) 
together with Eq.~(\ref{chiapprox}) gives the dispersion relation as a fourth
order polynomial in $\omega$ with real coefficients. Keeping all terms in 
this expression, which is valid for wavenumbers 
$\left|\hat{k}\right|\ll \Gamma_{\rm b}\beta_{\rm b}/\Theta$, 
far from resonance with the background protons, the low frequency modes
are shown in Fig.~\ref{modeplot}.

\section{Discussion}
From Fig.~\ref{modeplot} it can be seen that 
the $\varepsilon=1$ mode  (grey curves) resonates with the beam particles
when $k$ is close to their inverse gyro-radius, i.e., $\hat{k}\approx1$. 
This is the cyclotron resonant
mode considered by, amongst others, 
Pohl \& Schlickeiser \cite{pohlschlickeiser}. However, for low to moderate 
values of the background temperature, its growth rate is very much slower than
that of the non-resonant, current-driven mode (black curves). 
Near the peak of the unstable band of the non-resonant mode, the 
waves are strongly modified by the beam and do not resemble Alfv\'en waves:
the phase velocity of the 
non-resonant mode is lower than that of the 
resonant mode, and both speeds generally 
exceed $v_{\rm A}$ substantially. 
At very large wavelengths, the modes converge and become Alfv\'en waves.
This also happens for large $k$ values outside the unstable band,
but only if these do not resonate with the background protons. 
  
Eqs~(\ref{bell2}) and
(\ref{bell3}), show that in a cold background with $n_{\rm b}\approx
n_{\rm p}$, the non-resonant mode
has a maximum growth rate comparable to that of electrostatic 
oscillations, although at much shorter (parallel) wavelengths. 
Only a relatively modest temperature $\Theta_{\rm p}\approx\omega_{\rm
  c}/\omega_{\rm pp}$ suffices to damp such waves, as can be seen from
Eq.~(\ref{thermaleffects}). Physically, this can be understood by noting that,
as the background temperature rises, an increasing number of particles react
to the wave perturbation as if they were unmagnetized. This reduces the 
effective current driving the instability. Ultimately, if all particles react
in the same way to the perturbation 
(i.e., all are magnetized or all are unmagnetized) the plasma behaves as a
fully compensated system with no driving current.  
Eq.~(\ref{thermaleffects}) also shows that, for fixed
temperature, the growth rate scales with the cyclotron frequency of the 
background protons. For $\Theta_{\rm p}\approx1$ and $n_{\rm b}\approx n_{\rm
  p}$, the maximum growth rate is approximately equal to $\omega_{\rm c}$. 

The crucial question of the nonlinear evolution of the 
system has been discussed in the nonrelativistic case by  
Bell \cite{bell04} and Milosavljevi\'c \& Nakar \cite{nakar05}. 
Saturation can be expected when the currents associated 
with the growing waves become comparable to the current induced by the beam. 
Applying this to the relativistic case, one finds
for the amplitude $B_{\rm w}$ 
of the magnetic field of the perturbations
\eqb
\left|\vec{k}\wedge\vec{B}_{\rm w}\right| \approx q_{\rm b}n_{\rm b}c\beta
\eqe
If we assume that the largest wavelength in our system is of the order 
of the gyroradius of the beam protons, we can estimate the maximum
magnetic field energy density to be  
\eqb
{B_{\rm w}^2\over 8\pi} \approx \frac{1}{2}n_{\rm b} \Gamma m_{\rm p}c^2
\eqe

This suggests that the entire energy content of the beam is, at least 
initially, transferred into transverse field fluctuations. 
The efficiency of magnetic field generation by this mechanism
(i.e., the ratio of the saturated field energy density to that in the
incoming stream) is somewhat higher 
than that attributed to the Weibel instability in the case
of a relativistic shock in a pair plasma \cite{spitkovsky05}. The Weibel
instability is not thought to be effective in mediating shocks in
electron/proton plasmas \cite{lyubarskyeichler05}. However,
the efficacy of the instability described here depends strongly on the ability
of the plasma to convert the energy input at small scales into a larger scale
magnetic field.

\subsection*{Acknowledgements} This research was jointly supported by Cosmogrid 
and the Max-Planck-Institut f\"ur Kernphysik.


\section*{References}
\begin{thebibliography}{10}

\bibitem{achterberg83}
A. Achterberg, Astronomy \& Astrophysics {\bf 119}, 274 (1983)
\bibitem{bell04}
A.R. Bell, Mon.\ Not.\ R.\ Astron.\ Soc.\ {\bf 353}, 550 (2004)
\bibitem{bell05}
A.R. Bell, Mon.\ Not.\ R.\ Astron.\ Soc.\ {\bf 358}, 181 (2005)
\bibitem{belllucek}
A.R. Bell, S.G. Lucek, Mon.\ Not.\ R.\ Astron.\ Soc.\ {\bf 321}, 433 (2001)
\bibitem{blandfordeichler87}
R. Blandford, D. Eichler, Physics Reports, {\bf 154}, 1, 1 (1987)
\bibitem{drury83}
L. O'C. Drury, Rep. Prog. Phys., 46, 973 (1983)
\bibitem{hededalnishikawa}
C.B. Hededal, K.-I. Nishikawa, Astrophysical Journal, 623, L89 (2005)
\bibitem{jonesellison91}
F. Jones, D. Ellison, 58, 259, (1991) 
\bibitem{lyubarskyeichler05}
Y. Lyubarsky, D. Eichler, Astrophysical Journal, 647,1250 (2006)
\bibitem{mckenzievoelk82}
J.F. McKenzie, H.J. V\"olk, Astronomy \& Astrophysics {\bf 116}, 191 (1982)
\bibitem{medvedevloeb99}
M.V. Medvedev, A. Loeb, Astrophysical Journal, 526,697 (1999)
\bibitem{nakar05}
M. Milosavljevi\'c, E. Nakar, astro-ph/0512548 (2005)
\bibitem{nishikawaetal}
K.-I. Nishikawa, P. Hardee, G. Richardson, R. Preece, H. Sol, G.J. Fishman,
Astrophysical Journal, 595, 555 (2003)
\bibitem{grbreview}
T. Piran, Rev.Mod.Phys. {\bf 76}, 1143 (2004)
\bibitem{pohlschlickeiser}
M. Pohl, R. Schlickeiser, Astronomy \& Astrophysics {\bf 354}, 395 (2000)
\bibitem{lerchepohlschlickeiser}
M. Pohl, I. Lerche,  R. Schlickeiser, Astronomy \& Astrophysics {\bf 383}, 309 (2002)
\bibitem{silvaetal}
L.O. Silva, R.A. Fonseca, J.W. Tonge, J.M. Dawson, W.B. Mori, M.V. Medvedev,
Astrophysical Journal, 596, L121 (2003)
\bibitem{spitkovsky05}
A. Spitkovsky, 2005AIPC, 801, 345S, (2005)
\bibitem{wiersmaachterberg04}
J. Wiersma, A. Achterberg, Astronomy \& Astrophysics {\bf 428}, 365 (2004)
\bibitem{yoon90}
P.H. Yoon, Phys. Fluids B. {\bf 2}, 4, (1990)

\end{thebibliography}
\end{document}